\documentclass[%
 reprint,
 amsmath,amssymb,
 aps,
pra,
]{revtex4-1}

\usepackage{color}
\usepackage{graphicx}
\usepackage{dcolumn}
\usepackage{bm}
\usepackage{units}
\usepackage{epsfig, amsmath,amsfonts, amssymb,graphicx,amsthm,color}
\usepackage{upgreek} 

\begin{document}

\preprint{APS/123-QED}

\title{Laser stimulated deexcitation  of Rydberg antihydrogen atoms }

\author{D. Comparat}
\affiliation{Laboratoire Aim\'{e} Cotton, CNRS, Univ. Paris-Sud, ENS Paris Saclay, Universit\'e Paris-Saclay, B\^{a}t. 505, 91405 Orsay, France}

\author{C. Malbrunot}
\affiliation{Experimental Physics Department, CERN, Gen\`eve 23, 1211, Switzerland}

\date{\today}

\begin{abstract}
Antihydrogen atoms are routinely formed at CERN in a broad range of Rydberg states. Ground-state anti-atoms, those useful for precision measurements, are eventually produced through spontaneous decay. However given the long lifetime of Rydberg states the number of ground-state antihydrogen atoms usable is small, in particular for experiments relying on the production of a beam of antihydrogen atoms. Therefore, it is of high interest to efficiently stimulate the decay in order to retain a higher fraction of ground-state atoms for measurements. We propose a method that optimally mixes the high angular momentum states with low ones enabling to stimulate, using a broadband frequency laser, the deexcitation toward low-lying states, which then spontaneously decay to ground-state. We evaluated the method in realistic antihydrogen experimental conditions. For instance, starting with an initial distribution of atoms within the $n=20-30$ manifolds, as formed through charge exchange mechanism, we show that more than 80\% of  antihydrogen atoms will be deexcited to the ground-state within \unit[100]{ns} using a laser producing \unit[2]{J} at \unit[828]{nm}.
\end{abstract}

\maketitle

\section{Introduction}

Recent breakthroughs were achieved in spectroscopy measurements on antihydrogen atoms which led to stringent tests of the CPT symmetry, the combination of the charge conjugation, parity and time reversal symmetries
\cite{ALP182,ahmadi2016improved,2017Natur.548...66A}.
These measurements were all performed on magnetically trapped ground-state antihydrogen atoms which is currently the only method succeeding in accumulating enough ground-state atoms for measurements.
Indeed, antihydrogen atoms formed at CERN, using three-body recombination or charge exchange processes, are produced in a broad range of Rydberg states including all possible angular momentum states \cite{ROB08,RAD14}.
The highly excited atoms must first decay before a precision measurement can be performed. Trapped atoms can be held on for hours \cite{ALP172} so that the produced Rydberg atoms have ample time to decay. In beam experiments however, the spontaneous lifetimes of the Rydberg states are hindering a fast enough ground-state population \cite{MAL18}. 
Neglecting first the effect of external fields on the spontaneous lifetime of a ($n,l,m$) state (where $|m| \leq l < n$), the lifetime can be approximated by \cite{2003PhRvA..68c0502F}: 
\begin{equation*}
		\tau \approx \left( \frac{n}{30} \right)^3 \left( \frac{l+1/2}{30} \right)^2 \times \unit[2.4]{ms}.
		\end{equation*}

This result is also confirmed in a magnetic field environment.  For instance Ref.~\cite{Topccu2006} shows that within the \unit[1-5]{Tesla} field present in antihydrogen experiments and for the three-body recombination formation mechanism, only 10\% of the population with $n<30$ reach the ground-state in \unit[100]{$\mu$s}. 
A much longer time ($\sim$\unit[2]{ms}) is required to have 50\% of the population reaching the ground-state due to the large proportion of states with high angular momentum. Because antihydrogen atoms are typically formed with velocities of $\unit[1000]{ms^{-1}}$ (corresponding to the mean velocity of a Maxwell-Boltzmann distribution at $\sim$\unit[50]{K}), if they are not trapped, they will hit the walls of the formation apparatus and annihilate long before any spontaneous deexcitation to ground-state can occur.

\begin{figure}
\centering
\includegraphics[width=\linewidth]{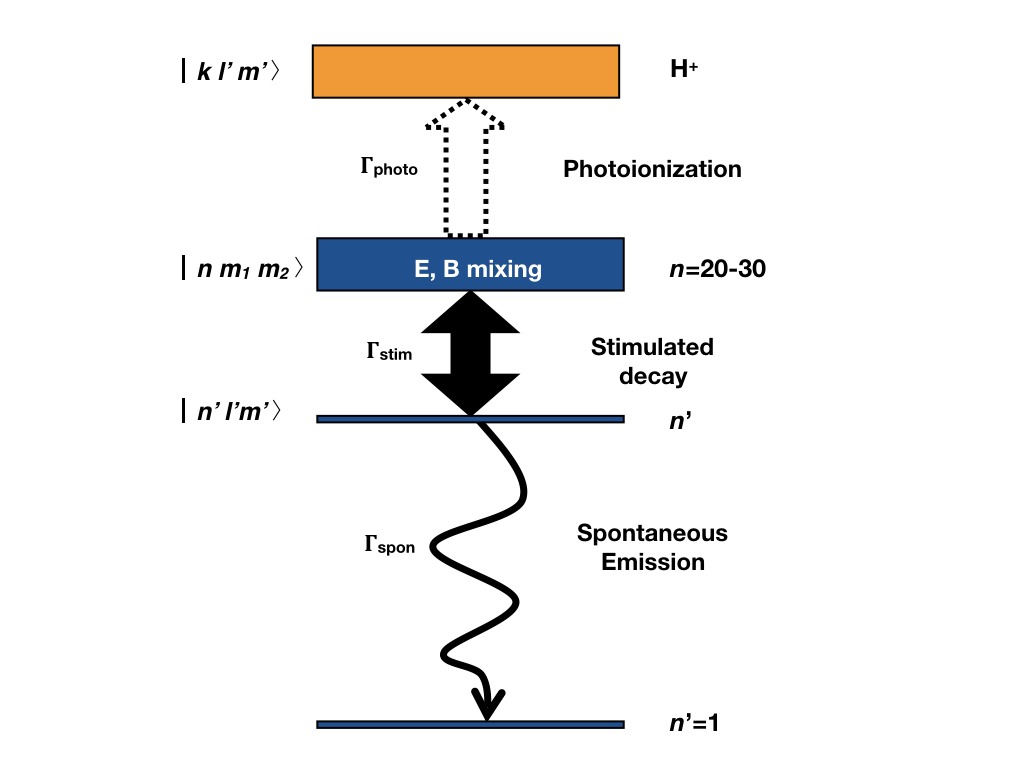}
\caption{Principle of the stimulated decay of Rydberg antihydrogen atoms. Combined electric $\bm E$ and magnetic $\bm B$ fields mix the Rydberg  states to form new $|n m_1 m_2\rangle$ states that can then be stimulated down to $n'$ at a rate $\Gamma_{\rm stim}$. This state will quickly decay toward the ground-state. The other basis states and the competing photoionization mechanism are shown.}
\label{fig:decay_principle}
\end{figure}

It is therefore of high interest to enhance the decay. It has been suggested that coupling or mixing high angular momentum states with low ones may accelerate the decay \cite{2011JPhB...44n5003H}. 
The main idea of the present article is to mix all angular momenta using an electric field added to the already present \unit[1-5]{Tesla} magnetic field of the antihydrogen experiments and use a laser to stimulate the decay from high-lying $n$ states to a deep-lying one with a short spontaneous lifetime. The principle of the proposed scheme is sketched in Fig.~\ref{fig:decay_principle}.

We first estimate the feasibility of the method  using a simple model assuming a fully mixed system and confirm that the laser power required is compatible with existing lasers and that photoionization can be drastically reduced by choosing a low enough $n'$ manifold. We then discuss the validity of the first order treatment in the electric and magnetic fields and show that an optimal configuration of fields can lead to a large mixing of the states. 
Finally using this optimal configuration of fields  we study in more details the effect of the laser power and polarization.

\section{General considerations assuming  fully mixed states}

For the sake of simplicity we focus here on antihydrogen  formed in  a pulsed charge exchange process \cite{2016PhRvA..94b2714K}.
The other formation mechanism used by the antihydrogen experiments is based on ``continuous'' three-body recombination of positrons and antiprotons (lasting as long as the antiproton and positron clouds can be kept in interaction, typically several hundreds of milli-seconds) producing typically $n>40$ 
\cite{ROB08,RAD14}.
This ``continuous'' production mode will be the focus of another publication where we will show that THz light can be efficiently used to stimulate the decay \cite{COM18}. Consequently, we assume an initial distribution of $n=20-30$ states with full degeneracy in $l$ and $m$ as planned to be produced in the AEgIS experiment \cite{2016PhRvA..94b2714K}.\\
\indent Before studying in detail the mixing mechanism in \S\ref{s:mixing} let's first assume the presence of an electric and magnetic field resulting in a perfect state mixing. We can already see that such hypothesis is likely to be valid by estimating the first order Zeeman and Stark effects (see \S\ref{s:firstorder})  which indicate that for $n\sim 20-30$ (with $r\sim n^2 a_0$) an electric field of the order of $F\sim$\unit[100]{V/cm} is sufficient to produce a Stark effect bigger than the spacing  between Zeeman sublevels leading to a strong mixing of the (Zeeman) $|n l m\rangle$ states as expected.\\
The assumption of perfect state mixing will allow us to evaluate the laser properties needed to stimulate the decay. The first obvious requirement is that the laser has to be broadband: in order to cover all Rydberg states $n=20-30$ we need a laser linewidth on the order of $\sim$\unit[2$\pi \times$ 5000]{GHz}. Note, that with such bandwidth we  can also cover  $n=25-\infty$.

\subsection{Fully mixed states hypothesis}
We assume a fully $l,m$ mixed initial state  $|\psi_n\rangle \approx \sum_{l m} 1/n |n l m\rangle $ that is coupled to the lower $n'$ manifold  thanks to a (spectrally Lorentzian) laser of FWHM $\Gamma_{\rm L} =  2\pi \times \unit[5000]{GHz}$  and of central wavelength $\lambda = 2\pi c /\omega$. We first consider an isotropic  polarization of the light. We can then calculate the  stimulated emission and photoionization rates  for a given laser intensity $I$.
		
		\subsubsection{Stimulated emission and photoionization rates}

The stimulated emission rate $\Gamma_{\rm stim}$   under an  unpolarized light is given by
the sum  over all polarizations $q$ and  over all final states $l' m'$: $\Gamma_{\rm stim} 
=  \frac{2 I e^2 a_0^2}{\hbar^2 \epsilon_0 c \Gamma_{\rm L}}
\frac{1}{3}	\sum_q \sum_{l' m'} \left| \langle \psi_n | r^{(q)}/a_0| n' l' m'  \rangle \right|^2
= \frac{2 I e^2 a_0^2}{\hbar^2 \epsilon_0 c \Gamma_{\rm L}}
\frac{1}{3 n^2}	 \sum_{l' l} \left| \langle n' l' \| r/a_0 \| n l \rangle \right|^2 
$, with
$
 \langle n' l'  \| r/a_0 \| n l  \rangle 
		 =   C_{l 0, 1 0}^{l' 0}   \sqrt{2 l +1}    { R}_{nl}^{n' l'} $ where $R_{nl}^{n'l'}$ is the radial
		  overlap  given 
		  in Eq. (\ref{radial_overlap}) of the appendix.
		 
Similarly, 
using the extra photon energy above the ionization threshold  given by
 $E=  k^2 Ry =\hbar \omega - Ry/n^2$, we find  the
 photoionization cross section:
$ \sigma_{n }^{k  } = \frac{1}{3} \sum_q \sum_{l'm'} 
4 \pi^2 \alpha a_0^2\left(\frac{1}{n^2} + k^2 \right) |  \langle k l' m'  | r^{(q)}/a_0 | \psi_n \rangle | ^2
=  \frac{1}{n^2}  \sum_{l l'} (2l+1) \sigma_{nl }^{k l'  }
$, where
  $\sigma_{nl }^{k l'  }$ is the photoionization cross section from 
$nl$ to $k l' $ given by:
$
\sigma_{nl }^{k l'  }  
= \frac{4 \pi^2 \alpha a_0^2}{3  }\left(\frac{1}{n^2} + k^2 \right) \frac{ \max(l,l')}{2l+1}  \left( R_{nl}^{k l'}\right)^2$. $ R_{nl}^{k l'}$ is the
radial overlap  given by Eqs. (\ref{radial_overlap_continuum_1}) and (\ref{radial_overlap_continuum_2}) in the appendix.

Finally, the photoionization rate is
$
\Gamma_{\rm photo} = \int d \tilde \omega  \frac{I(\tilde \omega)}{\hbar \tilde \omega} \sigma_{n }^{k(\tilde \omega)   }
$ and, because the cross section (and also the $\omega$ value) does not vary significantly over the laser spectral bandwidth we have
$
   \Gamma_{\rm photo} =  \frac{I}{\hbar \omega}  \sigma_{n }^{k  }
$

		\subsubsection{Saturation energy required}

\indent In order to have an efficient transfer toward the ground-state we need to transfer the population in a time scale compatible with  the laser pulse and  the spontaneous emission lifetime of the $n'$ levels. \\
We study two extreme cases. The first one assumes a short nanosecond (\unit[10]{ns}) laser pulse for which we calculate the saturated intensity $I=I_{\rm sat}(\unit[10]{ns})$ such that the stimulated decay rate is $\Gamma_{\rm stim} = 1/(\unit[10]{ns})$. In the second case we simply require a pulse duration comparable to the  spontaneous emission lifetime $\tau$ of the $n'$ manifold and calculate the saturated intensity $I_{\rm sat}(\tau)$ such that the stimulated decay rate is 1/$\tau$. Obviously $I_{\rm sat}(\unit[10]{ns}) =I_{\rm sat}(\tau) \frac{\tau}{\unit[10]{ns}}$, but we find useful to indicate both values. 
For this first study we choose $\tau^{-1} = \frac{1}{{n'}^2} \sum_{l'=0}^{n'-1} (2 l'+1) A_{n'l' } $ that is
the 
average decay rate (over all $l'm'$ levels) of the $n'$ manifold. 
It is calculated using  $A_{nl } = \sum_{n'=1}^{n-1} ( A_{nl }^{n' l+1  } + A_{nl }^{n' l-1  }) $ where $
A_{nl }^{n' l'  }
=  \frac{\alpha^4 c}{6 a_0} \left(\frac{1}{n'^2} - \frac{1}{n^2} \right)^3 \frac{\max(l,l')}{2l+1} \left(R_{nl}^{n' l'}\right)^2 $ is the  spontaneous emission rate from an $n l  $ state toward all $n' l'$ sub-levels ($\hbar \omega = \frac{Ry}{ n'^2} - \frac{Ry}{ n^2}$ is the transition energy).

	\begin{table}[h]		
		\begin{tabular}{|c||c|c|c|c|c|c|}
		\hline
	$n$ &	$n'$ & $\lambda$ (nm)  & $\tau$ (ns) & $E_{\rm sat}(\tau)$ (mJ) & $E_{\rm sat}$(\unit[10]{ns}) (mJ) & $\frac{\Gamma_{\rm stim}}{\Gamma_{\rm photo}} $ \\
		\hline
	30 &	10 & 10 250 & 1908 & 0.0043 & 0.82 &  0.78 \\
	20 &	10 & 12 150 & 1908 & 0.00029 & 0.056 &  0.78 \\
		\hline
		30 &	5 & 2343 & 86.5 & 4.4 & 38.5 &  6.0 \\
	20 &	5 & 2430 & 86.5 & 0.51 & 4.4 &  6.0 \\
						\hline
		30 &	4 & 1484 & 33 & 37 & 124 &  11.4 \\
	20 &	4 & 1518.8 & 33 & 4.5 & 14.9 &  11.4 \\
						\hline
		30 &	3 & 828.4 & 10 & 548 & 548 &  26.3 \\
	20 &	3 & 839.0 & 10 & 68.7 & 68.7 &  26.3 \\
						\hline
		30 &	2 & 366.1 & 2.1 & 20 708 & 4405 &  84.6 \\
	20 &	2 & 368.2 & 2.1 & 2670 & 568 &  84.6 \\
		\hline
	\end{tabular}
	\caption{Study of stimulated deexcitation. The laser linewidth $\Gamma_{\rm L}$ is assumed to be \unit[2$\pi \times$ 5000]{GHz}. The laser waist is \unit[1]{mm}.}
	\label{table_fully_mixed}
		\end{table}

\begin{figure}
	\centering
	\includegraphics[width=1\linewidth]{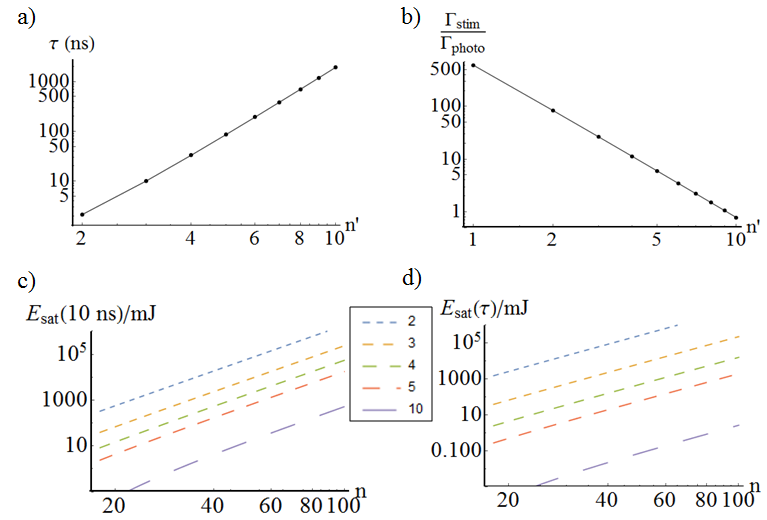}
	\caption{a)~:~average lifetime $\tau$ of the $n'$ manifold. b)~:~ratio of stimulated emission to photoionization rates.  c)~:~energy required to saturate the (fully mixed) $n \rightarrow n'$ transition~: $E_{\rm sat}$(\unit[10]{ns}) for an interaction time of \unit[10]{ns}. d)~:~$E_{\rm sat}(\tau)$ for an interaction time of $\tau$. As above the laser linewidth $\Gamma_{\rm L}$ is assumed to be \unit[2$\pi\times$5000]{GHz} and the waist \unit[1]{mm}. We find the approximate analytic formulas~:
	$\tau$ (ns)  $\approx 0.065 {n'}^{4.5}$, 
	$E_{\rm sat}(\tau)$ (mJ)$\approx 0.77 n^5 {n'}^{-9.5}$,
	$E_{\rm sat}(10 ns)$ (mJ) $\approx 0.007 {n'}^{-5.2} n^5$, 
	$\frac{\Gamma_{\rm stim}}{\Gamma_{\rm photo}}  \approx 630 {n'}^{-2.9}$.}
	\label{fig:mixed_plot}
\end{figure}

\indent 	 The results on the required saturation energy, 
assuming  a laser waist of \unit[1]{mm} (similar to the typical antiproton plasma size, see e.g. \cite{AGH18}),
as well as the ratio $\frac{\Gamma_{\rm stim}}{\Gamma_{\rm photo}} $, are shown in table \ref{table_fully_mixed} and in Fig. \ref{fig:mixed_plot}.

We first see that the  competing effect of the photoionization toward the continuum is relatively weak especially for low $n'$ values. This is due to the fact that the photoionization cross section drops quickly for high angular momentum states which are the most numerous states and that lower $n'$ implies shorter laser wavelength that also reduces the photoionization effect.\\
Obviously, the power required to deexcite increases with decreasing $n'$ due to the $n'^{3/2}$ scaling of the dipole matrix element and to the fact that fewer angular momenta ($|m'|\leq l' < n'$) exist and can be coupled to the initial ones.\\
\indent Based on these results we see that several laser choices are possible.
Toward $n'=10$, power in the Watt range is enough and some CO$_2$ lasers exist that can even allow a continuous deexcitation scheme for high $n$ values \cite{wan1985study,gupta1990various}.
But lower $n'$ are better to minimize photoionization. Deexcitation toward $n'=4$ seems feasible using a laser similar to the one used by AEgIS to excite positronium (a bound state of electron and positron) to Rydberg states \cite{AGH16}. However the use of a  nanosecond laser with a pulse much shorter than the spontaneous emission lifetime of the $n'$ states will limit the transfer  by equalizing population between upper and lower levels.
Therefore, targeting a lower state, like $n'=3$, will improve the deexcitation efficiency. Furthermore intense lasers that have a long pulse duration ($\sim$\unit[100]{ns}) in the Joule range exist at this wavelength. This is the case for the alexandrite \cite{WAL85} which can reach $\lambda >\unit[800]{nm}$ through heating of the medium \cite{KUP90}.

\section{Mixing in electric and magnetic fields}
\label{s:mixing}
For the next study we therefore restrict ourselves to $n'=3$, but most of the results will be valid for any other case.\\
Here we study in more detail the mixing produced by an electric and a magnetic field. We neglect the spins because Stark and Zeeman effects in the Rydberg $n$ manifold usually dominate the fine or hyperfine effects.

\subsection{First order Stark and Zeeman effects}
\label{s:firstorder}
Considering a given $n$ manifold, we have the following Hamiltonian: 
\begin{equation}\label{eq:hamiltonian}
 H = -\frac{1}{2 n^2} +{\bm r}. {\bm F} + \frac{1}{2} {\bm B} .{\bm L }  - \frac{1 }{8 } ( {\bm r} \times { \bm B})^2. 
\end{equation}
Atomic units will be assumed in this section.

Here,  we use perturbation theory in the field values (not to be confused with the standard state perturbation theory).
To the first order in the fields' values we simply have to deal with the perturbation
$ V_1 =  {\bm r}. {\bm F} +\frac{1}{2} {\bm B} . {\bm L }  $.
We can indeed  neglect the
 second order term since it is of the order of $n^6 F^2+n^4 B^2/4$ in atomic units (so for $B/(2.35 \times \unit[10^5]{T})$ and $F/(5.14 \times \unit[10^9]{V/cm})$) \cite{solovev1983second,PhysRevA.57.1149,braun1993discrete}. Thus, the second order becomes comparable to the first order (that is of the order of $ (B+3 n  F) n/2$) for $n\leq30$ and $F<\unit[1]{kV/cm} $ for a \unit[1-5]{Tesla} field. 
 Therefore, in our configuration, the first order should be accurate enough to extract the required fields values and laser energy.
 
The Hamiltonian (\ref{eq:hamiltonian}) has been studied by Pauli who showed that, for a given $n$ manifold,
$\bm r = - \frac{3}{2}  n {\bm A}$, where $\bm A$ is the Runge-Lenz vector. So in this manifold we can define
 new angular momenta
$\bm I_1=\frac{{\bm L} + {\bm A}}{2}$ and $\bm I_2=\frac{{\bm L} -{\bm A}}{2}$  that commute and verify
 $I_1=I_2= \frac{n-1}{2}$. We will use the 
$
|I_1 m_1 \rangle \otimes |I_2 m_2 \rangle  $ basis where the eigenvalues $m_1,m_2$ (on a given axis) take the values $-(n-1)/2,-(n-3)/2, . . . ,(n-1)/2$.
We define $\bm \omega_1=\frac{{{\bm B}}  - 3 n {\bm F} }{2}$ and $\bm \omega_2=\frac{{{\bm B}}  + 3 n {\bm F} }{2}$ such that
$V_1 = \bm \omega_1 . \bm I_1 + \bm \omega_2 . \bm I_2$.
That is trivial to diagonalize using the
$ |I_1 m_1 \rangle_{\bm  \omega_1}  |I_2 m_2 \rangle_{\bm  \omega_2} =  |n m_1 m_2 \rangle$  basis. This notation indicates that  $m_1$ is the projection of $I_1$ on the $\bm \omega_1$ axis. So the first order perturbation theory gives
\begin{equation}
\Delta  E^{(1)}_{m_1 m_2} =   \omega_1 m_1 +  \omega_2 m_2.
\end{equation}
where $\omega_i = \| \bm \omega_i \|$.

For a pure electric field we restore  the pure Stark effect
$\Delta  E^{(1)}_{m_1 m2} =   \frac{3 }{2}  n  F (m_1 +  m_2) $ with a clear relation to the
 parabolic basis  $|n, k,m\rangle$ linked to  $\hat H,\hat A_z, \hat L_z$ eigenvalues: $k=-(m_1+m_2) $ and $m= m_2-m_1$ \cite{demkov1970energy,solovev1983second}.

In a pure magnetic field we  restore the standard Zeeman shift
$\Delta  E^{(1)}_{m_1 m2} =   \frac{1 }{2}   B (m_1 +  m_2) $
where $m_1+m_2 = m$ because $\bm I_1 + \bm I_2= {\bm L} $. 

Using $ {\bm L}  = \bm I_1 + \bm I_2 $ and the standard sum of the two angular momenta leads to 
$|n l m\rangle_{\bm B} = \sum_{m_1 m_2} 
C_{I_1 m_1, I_2 m_2}^{l m}
|I_1 m_1 \rangle_{\bm B} |I_2 m_2 \rangle_{\bm B} $ or
$
|I_1 m_1 \rangle_{\bm B} |I_2 m_2 \rangle_{\bm B}  = \sum_{m l} 
C_{I_1 m_1, I_2 m_2}^{l m}  |n l m\rangle_{\bm B} = \sum_{l} 
C_{I_1 m_1, I_2 m_2}^{l, m_1+m_2}  |n l ,m=m_1+m_2\rangle_{\bm B}
$
where the subscript ${}_{\bm B}$ indicates that the quantization axis   $z$ is along $\bm B$.

If we define $\alpha_1$ and $\alpha_2$ as the angles between the magnetic field $\bm B$ axis and the vectors ${\bm \omega}_{1}$ and ${\bm \omega}_{2}$ respectively by the use of the Wigner $D$ rotation matrix, we have \cite{demkov1970energy} (Eq. 1.4 (35) of \cite{varshalovich1987quantum})~:

$$|I_1 m_1 \rangle_{\bm \omega_1} = \sum_{m'_1 = - I_1}^{m'_1 = I_1} D_{m_1, m'_1}^{I_1} (0,\alpha_1,0)
|I_1 m'_1 \rangle_{\bm B}.$$
 
So by combining the equations (we use real  Clesch-Gordan and real  Wigner (small) $d$-matrix $  d_{m_2, m'_2}^{I_2} (\alpha_2) = D_{m_2, m'_2}^{I_2} (0,\alpha_2,0)$) we get:
\begin{widetext}
\begin{eqnarray}
	 |n m_1 m_2 \rangle_{\bm \omega_1, \bm \omega_2} & \equiv & |I_1 m_1 \rangle_{\bm \omega_1} |I_2 m_2 \rangle_{\bm \omega_2} =
\sum_{ l=0}^{n-1} \sum_{m=-l}^l  \langle n l m  | n m_1 m_2 \rangle
|n l m \rangle_{\bm B} \\
  \langle n l m  | n m_1 m_2 \rangle & \equiv &  \sum_{m'_1 = - I_1}^{m'_1 = I_1 = \frac{n-1}{2}} \sum_{m'_2 = - I_2}^{m'_2 = I_2 = \frac{n-1}{2}}  d_{m_1, m'_1}^{I_1} (\alpha_1^{(n)} )  d_{m_2, m'_2}^{I_2} (\alpha_2^{(n)})
C_{I_1 m_1', I_2 m_2'}^{l m},  \label{decomposition_parabolic}
\end{eqnarray}
\end{widetext}
where we have written $\alpha_1^{(n)}$ to stress that the angles depend on $n$ and not only on the fields' values.
Similarly
$ |n l m \rangle_{\bm B}  =
\sum_{ m_1,m_2}  \langle n m_1 m_2 |  n l m  \rangle |n m_1 m_2 \rangle_{\bm \omega_1, \bm \omega_2}$.
And we have  the closure expression  ${\bm 1}_n =  \sum_{m_1,  m_2}
  |  n   m_1  m_2 \rangle  \langle   n   m_1  m_2 |  = \sum_{l m}   |  n   l m  \rangle_{\bm B}  {}_{\bm B}\langle   n   l m  |$.
  
  We can note that using such a ($ m_1, m_2 $) formalism the  states will always be given in a parabolic basis not in a spherical basis, so even in a pure magnetic field the eigenstates $|n m_1 m_2 \rangle$  will not correspond to the $|nl m\rangle$ ones: $l$ is mixed in the $|n m_1 m_2 \rangle$ basis.

\subsection{Laser transitions under  combined electric and magnetic fields}

In order to choose the electric field which most optimally mixes the states, we need to calculate all transitions dipole moments from a given $|n m_1 m_2\rangle$ toward each states of the $n'$ manifold. The Zeeman and Stark effect being small for $n'=3$ we use the $|n' l' m' \rangle$ basis. So the
stimulated emission rate, for a $q=\pm 1,0$ polarization, from an $ | n m_1 m_2 \rangle$ state toward all the $ n'$ manifolds is given by :
\begin{eqnarray*}
\lefteqn{\Gamma_{\rm stim}^{n  m_1 m_2;  n' ; q}  = 
 \frac{2 I e^2 a_0^2}{\hbar^2 \epsilon_0 c \Gamma_{\rm L}}
\sum_{ l', m'} \left| \langle  n'  l' m' |  \frac{r^{(q)}}{a_0} | n m_1 m_2 \rangle \right|^2 } \\
&  = & \frac{2 I e^2 a_0^2}{\hbar^2 \epsilon_0 c \Gamma_{\rm L}}
   \sum_{m,  l' } \left| \sum_{l= l' \pm 1}   \langle  n'  l' \, m+q  | \frac{r^{(q)}}{a_0} | n l m \rangle \langle n l m  | n m_1 m_2 \rangle  \right|^2 
\end{eqnarray*}
that can be calculated using Eq.~(\ref{dipole}) of the appendix and (\ref{decomposition_parabolic}) in \ref{s:firstorder}.

The laser driven evolution of these states may be quite complex with $n^2$ levels coupled to $n'^2$ ones (and to the continuum).
 Rate equations is sufficient to treat the problem assuming
that the broadband  laser used to stimulate the deexcitation is incoherent (which is likely to be the case).
 Furthermore, because we have chosen the $n'=3$ with a fast spontaneous emission lifetime, the population of the $n'=3$ levels will be small and the re-excitation process from $n'=3$ to the $n$ manifold will not be severe. We can therefore consider that the  $n  m_1 m_2$ levels are isolated from each other and thus simplify the picture to a 4 levels rate equation system as shown in Fig.~\ref{fig:decay_principle}:
 \begin{eqnarray}
 \frac{d N_{\rm H^+}}{d t } & = &\Gamma_{\rm photo} N_{ n} \label{rate_eq} \\
  \frac{d N_{n }}{d t } & = & - \left( \Gamma_{\rm stim}+ \Gamma_{\rm photo} \right)  N_{ n } +  \Gamma_{\rm stim} N_{ n '} \nonumber \\
    \frac{d N_{ n' }}{d t } & = &  \Gamma_{\rm stim} (N_{ n } - N_{ n'}) -  \Gamma_{\rm spon} N_{ n' } \nonumber \\
    \frac{d N_{ 1 }}{d t } & = &   \Gamma_{\rm spon} N_{ n' } \nonumber
\end{eqnarray}

Cumbersome analytical solution exists for $P_{n  m_1 m_2 } ^{ (q) }(t)$,
the transfer of a given $|n m_1 m_2\rangle$ state to ground-state,
and we use them throughout this article. However in this section we can simplify the solution because,
as seen in  table \ref{table_fully_mixed} and Fig. \ref{fig:mixed_plot}) we can safely  neglect the photoionisation. We can also assume an instantaneous spontaneous emission from $n'=3$ to ground-state if the laser pulse duration is  much longer than the spontaneous emission lifetime of \unit[10]{ns} (which can be considered the case for a e.g.~\unit[100]{ns}-long laser pulse as is the case for the alexandrite for example). In this case, the model becomes a simple two levels model, and the transfer of a given $|n m_1 m_2\rangle$ state to ground-state, after the application of the laser of polarization $q$ and of duration $t$, is  given by 
$
P_{n  m_1 m_2} ^{  (q)} (t) = 1 - e^{-t \Gamma_{\rm stim}^{n  m_1 m_2;  n' ; q}}
$

\begin{figure}
	\centering
	\includegraphics[width=1\linewidth]{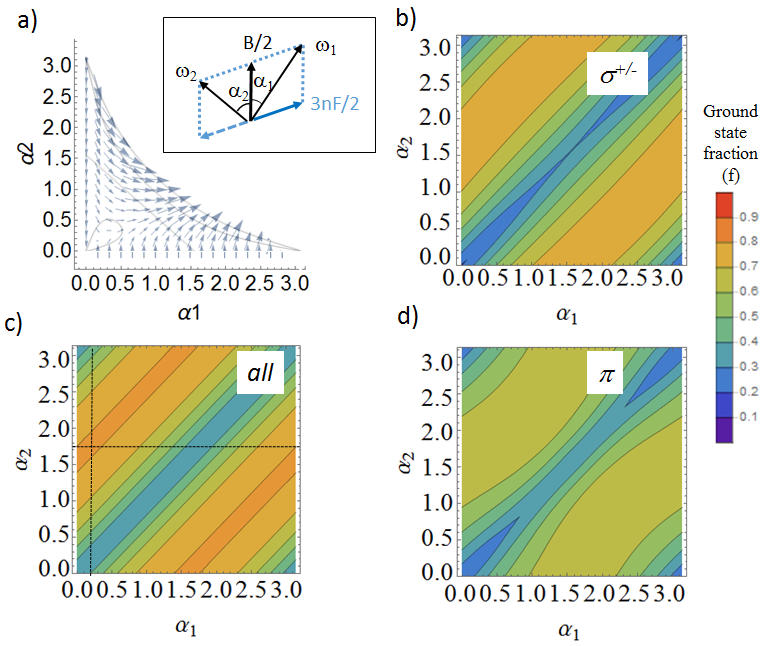}
	\caption{a)~In the inset :  definition of the $\alpha_1$ and $\alpha_2$ angles. Schematic of the electric $\bm F$ field orientation with respect to the magnetic $\bm B$ field. The frontiers where the length of the $3 n \bm F/ B$ vector (in atomic units) is 0.5, 1, 1.5 and 2 are shown and no values larger than 2 is plotted. b), c) and d)~:~Fraction $(f)$ of the initial atoms reaching ground-state as a function of the mixing induced by an electric field $\bm F$ for the $n=20 \rightarrow n'=3$ case and for a laser of
$\sim$\unit[150]{mJ} energy with a pulse duration of \unit[100]{ns}, a linewidth $\Gamma_{\rm L}=$\unit[2$\pi\times$5000]{GHz} and a waist of \unit[1]{mm}. For
	a circular polarized laser b), for an isotropic laser polarization c) and for a $\pi$ polarized laser (linear and parallel  polarization compared to the $\bm B$ axis) d). The dashed lines in c) indicate a judicious choice of $\alpha_1$ and $\alpha_2$ to achieve maximal mixing and thus maximal ground-state population.}
	\label{fig:angles_effect}
\end{figure}

 Assuming an equidistribution of the 
 initial $|n m_1 m_2\rangle$  states  we sum over these $n^2$ states to get
 the total amount of atoms reaching ground-state. Results 
are given in Fig.~\ref{fig:angles_effect} as a function of $\alpha_1$ and $\alpha_2$ for different laser polarizations. We have used, in cartesian coordinates with $\bm B=(0,0,B)$, the result
$ \frac{3 n \bm F}{B} = (0, 2 \frac{ \sin(\alpha_1) \sin(\alpha_2) }{\sin(\alpha_1+\alpha_2)} 
, 2 \frac{ \sin(\alpha_1-\alpha_2) }{\sin(\alpha_1+\alpha_2)}
)$.\\
\indent The calculation has been performed for $n=20$  with a laser energy of $\sim$\unit[150]{mJ} and a pulse of \unit[100]{ns} (but similar results hold for other $n$ states or other laser power values). 

The first important result is that it is possible to efficiently mix the states by adding an electric field to a magnetic field validating the assumption taken in the previous section.
The transfer efficiency is very high for several values and orientations of the electric field. As expected, an un-favourable configuration is that where the fields are orthogonal to each other or in the case of a too small electric field. A typical favourable configuration is when the electric field axis is oriented with a small angle with respect to the magnetic field axis and has a value such that $3nF\sim B$ in atomic units. So for instance, for $n=30$, in a \unit[1]{Tesla} magnetic field, a \unit[280]{V/cm} electric field with \unit[160]{degree} angle (corresponding  to $\alpha_1=0.187$, $\alpha_2 = 1.777$) is a good choice to efficiently mix the states as shown by the dashed lines in Fig.~\ref{fig:angles_effect}~c).
These magnetic and electric field values are small enough to allow the use of the first order perturbation theory approach.  This is  indicated by the contour of the $3 n  F/B$ values in Fig.~\ref{fig:angles_effect} a) that should be smaller than $\sim 2$ to avoid large second order effects.

\section{Decay in the optimized fields configuration}

We now choose the values of $\alpha_1$ , $\alpha_2$ marked in Fig.~\ref{fig:angles_effect}~c) to study more precisely the deexcitation mechanism.
Before doing so we stress that the choice of the initial states created by the addition of the electric field is not obvious.
It is beyond the scope of this letter to study in detail the dynamical behaviour of the states mixing  during the application of the electric field. We can nevertheless mention that, in order to fully mix the levels, it should not be  switched on too fast (meaning in a fully diabatic manner).  In the case of 
our real and
non-oscillating Hamiltonian, the adiabaticity criterion to stay in the eigenstates $|n\rangle = |n m_1 m_2 \rangle_{\bm  \omega_1, \bm  \omega_2 } $, is the standard criterion (with simplified obvious notations) $ \sum_{m \neq n}  \frac{\hbar |\langle m |d V_1/dt |  n \rangle| }{  |E_n -E_m |^2 } $ \cite{Comparat2009}. It  can be estimated using  simple classical
vector arguments \cite{lutwak1997circular}:
the rotation rate of $\bm \omega_{i}$  must always be
slow compared to the precession rate $\bm \omega_{i}.\bm I_{i}/\hbar$  (for $i=1,2$). Such estimation leads to a rate in the high range of V/cm  per ns. Thus a rising time less than $\sim \unit[10-100]{ns}$ should be safe to ensure adiabaticity.
Because initially antihydrogen are formed in all $|n m_1 m_2 \rangle_{\bm  B} $ states we will assume adiabaticity and so an equidistribution of the 
$|n m_1 m_2 \rangle_{\bm  \omega_1, \bm  \omega_2 } $ states before the application of laser deexcitation.

\begin{figure}
\centering
\includegraphics[width=1\linewidth]{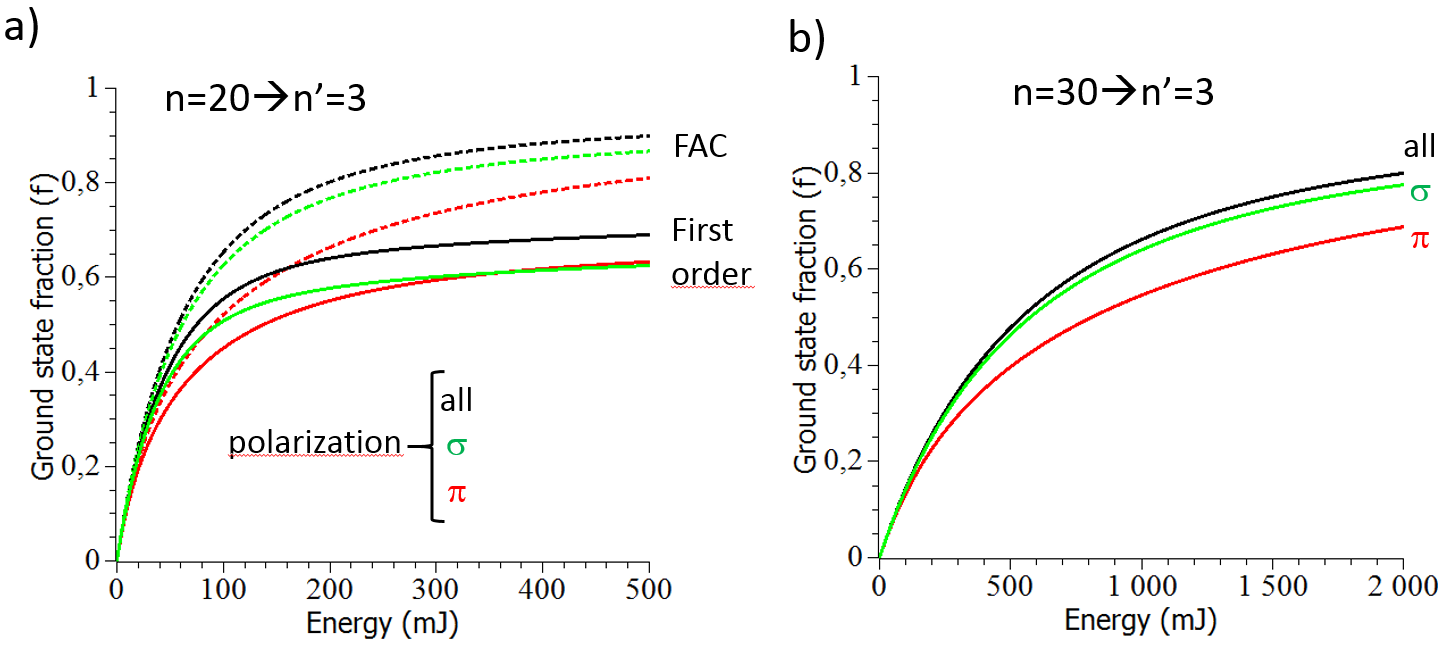}
\caption{ Fraction $(f)$ of the initial atoms reaching the ground-state from the $n=20$ a) or $n=30$ b) manifolds after stimulated deexcitation toward $n'=3$ in a \unit[1]{Tesla} magnetic field and a \unit[280]{V/cm} electric field with \unit[160]{degree} respective angle. Different laser polarizations are considered, see also Fig.\ref{fig:angles_effect} (black for isotropic, red for $\pi$ and green for $\sigma^{\pm}$). The laser has a \unit[5000]{GHz} linewidth (centered on the $n \rightarrow n'=3$ transition) and a waist of \unit[1]{mm}. In a) we added the comparison between our calculations using first order theory (solid lines) and the output of the Flexible Atomic Code (FAC) (dashed lines).}
\label{fig:decay_simulation}
\end{figure}

As before, we assume an abrupt application of a laser of duration $t=\unit[100]{ns}$.
The total population transfer toward the ground-state is plotted in Fig.~\ref{fig:decay_simulation} as a function of the laser power for two initial $n$-manifold ($n=20$ and $n=30$) and 3 laser polarizations:  $\sigma^{\pm},\pi$ or isotropic.
More precisely the total population transfer is calculated as the sum of the population transfer of the initially populated $|n  m_1 m_2  \rangle$ levels:
$ \frac{1}{n^2} \sum_{m_1,m_2} P_{n  m_1 m_2 } ^{  (q)} (t) $.  $P_{n  m_1 m_2 } ^{  (q)} (t) $ being calculated using equations (\ref{rate_eq}) with  the  stimulated decay rate $ \Gamma_{\rm stim}^{n  m_1 m_2;  n' ; q} $, the  photoionization  rate of the $n  m_1 m_2$ levels to the states of energy $k$ in the continuum (determined by the laser central wavelength)  $\Gamma_{\rm photo}$ and   the  spontaneous emission rate $\Gamma_{\rm spon} = 1/\tau$.

For $n=30$ to $n'=3$, an efficient transfer of nearly $80\%$ is obtained for a \unit[2]{J} laser energy. 
For the $n=20$ case the transfer requires, as expected, less laser energy  because of the larger transition dipole moment. But despite the fact that the field configuration is not optimal for $n=20$ (in that case we would have to choose $\alpha_1=0.151$ and $\alpha_2 = 0.756$), we still obtain around $60\%$ transfer.
In order to study the validity of our assumptions we perform the exact diagonalization of the Hamiltonian using the Flexible Atomic Code (FAC: a software package for the calculation of various atomic processes \cite{gu2008flexible}) for the same field configuration and including all states in the $n=1,3,20$ manifolds as well as the photoionization to the continuum.  The code
 takes into account the fine structure and the full Stark and Zeeman effects (including the quadratic Zeeman effect).
 We compare in Fig.~\ref{fig:decay_simulation}~a) those results to our first order calculation. A relatively good agreement is found. Interestingly, FAC predicts an even better mixing (which might be due to the quadratic Zeeman effect) showing that our estimated deexcitation efficiencies are probably slightly under-estimated.

\section{Conclusion}
We find that adding an electric field to the magnetic field already present in the antihydrogen apparatus will allow to very efficiently stimulate the decay of Rydberg states toward the ground-state. A continuous deexcitation might be possible using an intense CO$_2$ laser (that could also be used to directly create the antihydrogen atoms through stimulated radiative recombination \cite{wolf1993laser,muller1997production,amoretti2006search}).
A pulsed deexcitation toward $n'=3$ can be achieved using a heated alexandrite laser or toward $n'=4$ using an amplified OPG laser.
For narrower $n$ distribution, the laser power requirement will be more favorable because a smaller bandwidth will be required and thus less laser power would be needed to drive the transitions; the photoionization would also be reduced in the same ratio.
In order to further reduce the laser power we could also consider modifying the electric field strength during the laser pulse in order to more efficiently mix all levels. It may also be possible to mix the states using a radio-frequency field resonant to $m\rightarrow m\pm 1$ magnetic transitions or to add THz sources to stimulate the transitions $n \rightarrow n-1$ between Rydberg states \cite{COM18}.

Once in ground-state the atoms can be manipulated and reexcited to a well-defined state for targeted manipulations. The presented method will thus be very useful in mechanisms which envision the creation of an intense antihydrogen beam via well-controlled Stark acceleration \cite{AEG07} or magnetic focusing \cite{ASA05}. It can also allow a better trapping efficiency if a subsequent excitation is done in a state with a high magnetic moment (this will be the subject of an upcoming publication \cite{COM18}).
We therefore think that our proposal opens exciting future prospects to enhance the production of useful antihydrogen atoms.

\newpage

\section{Appendix}
\label{appendix}

We find useful to recall here some basic formulas to calculate  hydrogen properties especially because overlap of the radial wavefunctions of hydrogen, if  well known \cite{gordon1929berechnung}, are often misprint in  several articles (Eq. (25-27) of \cite{burgess1965tables}, (2.34) of \cite{storey1991fast} and (27) of \cite{gordon1929berechnung}) and textbook (Eq. (3.17) of \cite{bethe2012quantum}).

\subsection{Radial wavefunction}
We  use calligraphic notation (${\cal R}_{nl}$ or ${\cal R}_{El}$) for SI units and usual typography ($R_{nl}$ or  $R_{El}$) for atomic units.

The wavefunction of hydrogen atoms for bound states of energy $E=-\frac{Ry}{n^2}$ where $Ry = h c R_{\infty}  $ ($R_{\infty} = \frac{m_e c \alpha^2}{2 h})$  is the Rydberg energy,  is given by $\psi_{nlm}(r,\theta, \phi) = {\cal R}_{nl}(r) Y_{lm} (\theta, \phi)$
with $ {\cal R}_{nl}(r) = \frac{1}{a_0^{3/2}}  R_{nl}(r/a_0) $, $\rho = r/a_0$ and
\begin{eqnarray}
 R_{nl}(\rho ) & =& \frac{ 2}{n^2} e^{-\frac{\rho }{n}} \sqrt{\frac{(n-l-1)!}{ (l+n)!}} \left(\frac{2 \rho }{n}\right)^l L_{n-l-1}^{2 l+1}\left(\frac{2 \rho }{n}\right)  \nonumber \\
 & = & \frac{2}{n^2 (2 l+1)!}  e^{-\frac{\rho }{n}} \sqrt{\frac{(l+n)!}{ (n-l-1)!}} \left(\frac{2 \rho }{n}\right)^l \times \nonumber \\
 & & 
	_1F_1\left(l-n+1;2 l+2;\frac{2 \rho }{n}\right).
\end{eqnarray}

The functions are normalized : $1= \int_0^\infty R_{nl}(r)^2 r^2 dr =
 \int_0^\infty {\cal R}_{nl}(\rho)^2 \rho^2 d \rho$.

For continuum states
 $\psi_{Elm}(r,\theta, \phi) = {\cal R}_{E l}(r) Y_{lm} (\theta, \phi)$
 with energy $E=k^2 Ry $.\\
 Several normalizations (in energy, wavenumber, $k^2$,...) are possible \cite{landau1977quantum}. We choose here 
 the energy normalization :\\
$\delta(E-E') = \int_0^\infty {\cal R}_{El}(r) r^2 {\cal R}_{E' l}(r) dr $.
So with  
 ${\cal R}_{E l}(r) = \frac{1}{Ry^{1/2} a_0^{3/2}} R_{k l} (r/a_0)$ we have  the normalization through
 $\delta(k^2-k'^2) = \int_0^\infty R_{kl}(\rho) \rho^2 R_{k' l}(\rho) dr $ (so with a factor $\pi$ different compared to Ref. \cite{burgess1965tables}). We have (up to a phase factor)
 \begin{eqnarray}
 R_{kl}(\rho) &= &\frac{1 }{(2 l+1)!} e^{i k \rho } \sqrt{\frac{
 	2	\prod_{s=0}^{l} (1+s^2 k^2)  }
 			{1-e^{-\frac{2 \pi }{k}}}}  { (2  \rho) }^l \times \nonumber \\
 & & 
 	_1F_1\left(l-\frac{i}{k}+1;2 l+2;-2 i k \rho \right).
 \end{eqnarray}
 
The wavefunctions are similar to the bound state ones (through the modification $n \rightarrow i k$ due to the energy definition)
\cite{landau1977quantum,blaive2009comparison}.

\subsection{Reduced dipole matrix element}

 The Wigner-Eckart theorem indicates that  (bound-bound or bound-continuum) dipole $\bm d= e \bm r$ matrix elements  between $|n l m\rangle$ and $|n' l' m' \rangle$ states (or with $k$ in place of $n'$ for continuum states) are given by:
\begin{eqnarray}
	\langle n' l' m' | r^{(q)}/a_0 |n l m \rangle & =&  C_{l m, 1 q}^{l' m'} \frac{ \langle n' l'  \| r/a_0 \| n l  \rangle }{\sqrt{2 l' +1}} \label{dipole}  \\
		& = & C_{l m, 1 q}^{l' m'} C_{l 0, 1 0}^{l' 0} \frac{   \sqrt{2 l +1}   }{\sqrt{2 l' +1}} { R}_{nl}^{n' l'}. \nonumber
	\end{eqnarray}
The overlap 
$ R_{nl}^{n' l'} =  \int_0^\infty R_{nl}(\rho) \rho R_{n' l'}(\rho)  \rho^2 d\rho  = {\cal R}_{nl}^{n' l'}/a_0  $  is directly the atomic unit value:

\begin{widetext}
\begin{eqnarray}
R_{nl}^{n' l-1} & =& \frac{(-1)^{n'-l}}{4 (2 l-1)!}
\sqrt{\frac{(l+n)! \left(l+n'-1\right)!}{(-l+n-1)! \left(n'-l\right)!}}
\frac{(4 n n')^{l+1} (n-n')^{n+n'-2l-2} }{(n+n')^{n+n'}} \label{radial_overlap} \\
& &
\left(
\, _2F_1\left(l-n+1,l-n',2 l,-\frac{4 n n'}{\left(n-n'\right)^2}\right)-\frac{\left(n-n'\right)^2}{\left(n'+n\right)^2} \,
	_2F_1\left(l-n-1,l-n',2 l,-\frac{4 n n'}{\left(n-n'\right)^2}\right)
\right). \nonumber
\end{eqnarray}
\end{widetext}

One common case is when 
the atoms are evenly distributed among all $2l + 1$
possible $m$ initial states.
Using the sum rule $ \sum_{m q} 	|\langle n' l' m' | r^{(q)} |n l m \rangle|^2  = \frac{|\langle n' l'  \| r \| n l  \rangle| ^2}{2 l' +1} $ we see that for unpolarized light (intensity 1/3 for $\sigma^+$,   1/3 for $\pi$, 1/3 for $\sigma^-$ transition) the 
probability transition is independent on the initial state.
Another useful sum rule is 
$ \sum_{m' q} 	|\langle n' l' m' | r^{(q)} |n l m \rangle|^2  = \frac{|\langle n' l'  \| r \| n l  \rangle| ^2}{2 l +1} $.

\subsection{Spontaneous system}
The spontaneous emission rate between 
an $n l m $ and $n' l' m'$ level with photon angular frequency $\omega$ is given by
\begin{equation}
    A_{nl m}^{n' l' m } = \frac{\omega^3 e^2 a_0^2}{3 \pi \varepsilon_0 \hbar c^3} 	|\langle n' l' m' | r^{(q=m'-m)}/a_0 |n l m \rangle|^2. 
    \label{spontaneous}
\end{equation}
So when summed over the final states the spontaneous emission rate between 
an $n l m $ level and (all) $n' l' $ levels is given by
$A_{nl m}^{n' l'  } = \frac{e^2 a_0^2 \omega^3}{3 \pi \varepsilon_0 \hbar c^3} 	\frac{|\langle n' l'  \| r/a_0 \| n l  \rangle| ^2}{2 l +1} $ that does not depend on $m$ and can be noted $A_{nl }^{n' l'  }$ the  spontaneous emission rate from an $n l  $ toward all $n' l'$ levels. 
 Using $\hbar \omega = \frac{Ry}{ n'^2} - \frac{Ry}{ n^2}$ we find:

\begin{eqnarray}
A_{nl }^{n' l'  } &=& \frac{e^2 \omega^3}{3 \pi \varepsilon_0 \hbar c^3} 	\frac{\max(l,l')}{2 l +1} ( {\cal R}_{ n' l'}^{n l})^2 \\
&= & \frac{\alpha^4 c}{6 a_0} \left(\frac{1}{n'^2} - \frac{1}{n^2} \right)^3 \frac{\max(l,l')}{2l+1} \left(R_{nl}^{n' l'}\right)^2  \nonumber \\
 & = &  \frac{e^2 a_0^2 \omega^3}{3 \pi \varepsilon_0 \hbar c^3} 	\frac{|\langle n' l'  \| r/a_0 \| n l  \rangle| ^2}{2 l +1} \nonumber \\
 &=&
\frac{A_{nl m}^{n' l' m } }{
 \left( C_{l m, 1 q=m'-m}^{l' m'} \right)^2 \frac{2l+1}{2l'+1}
} = \frac{A_{nl m}^{n' l' m' } }{
\left( C_{l' m', 1 (-q)=(m-m')}^{l m} \right)^2 
}.  \nonumber
\end{eqnarray}

\subsection{Stimulated emission rate $\Gamma'$}

The stimulated emission rate $\Gamma'$ between 
an excited  $n l m $ and $n' l' m'$ level can be calculated in the same way.
The general formula for 2 levels $e$ and $g$ (separated in energy by $\hbar \omega_{\rm eg}$ , a dipole $d_{\rm e g} = e	a_0 \langle n' l' m' | r^{(q=m-m')}/a_0 |n l m \rangle$ transition
and a 
 laser polarization vector $\bm \epsilon$) is
 $$\Gamma'  =  \frac{\int L(\omega)   I(\omega) {\rm d}\omega  \uppi |{\bm d}_{\rm e g} . {\bm \epsilon}|^2  }{\hbar^2 \epsilon_0 c},$$
 where
$L(\omega)
= \frac{1}{ \uppi} \frac{\Gamma/2}{(\omega-\omega_{\rm eg})^2 + (\Gamma/2)^2} $ is the Lorentzian spectral spectrum for the spontaneous emission and $I (\omega)$ is the laser irradiance spectral distribution (throughout the article we use improperly the word intensity).

 For example for a Lorentzian spectrum $I (\omega)  =  \frac{I}{\uppi} \frac{\Gamma_{\rm L} /2 }{(\omega-\omega_0)^2 + (\Gamma_{\rm L} /2 )^2}$ ($I = \int I(\omega) d \omega $  is the full laser irradiance so the laser electric field is $E=\sqrt{2 I/\epsilon_0 c}$), we have $\Gamma'=  \frac{\Omega^2/2}{(\omega_0-\omega_{\rm eg})^2 + ((\Gamma_{\rm L}+\Gamma)/2)^2}  \frac{\Gamma_{\rm L}+\Gamma }{2}  
$, where $\Omega = \bf d_{\rm e g}.\bf E/\hbar$ is the Rabi frequency.

For a broadband laser, where $\Gamma_{\rm L} \gg \Gamma$, the final rate at resonance is
\begin{equation}
    \Gamma'=  \frac{\Omega^2}{\Gamma_{\rm L}}   =  \frac{2 I  |d_{\rm e g} \epsilon_q|^2 }{\hbar^2 \epsilon_0 c \Gamma_{\rm L}}.
    \label{gamma_stim}
\end{equation}

Therefore an average rate for a fully mixed state $ |\psi_n\rangle \approx \sum_{l m} 1/n |n l m\rangle $ under an isotropic (unpolarized) light is given by
the sum  over all $l' m'$ transition rates and so is $\Gamma' 
=  \frac{2 I e^2 a_0^2}{\hbar^2 \epsilon_0 c \Gamma_{\rm L}}
\frac{1}{3}	\sum_q \sum_{l' m'} \left| \langle \psi_n | r^{(q)}/a_0| n' l' m'  \rangle \right|^2
= \frac{2 I e^2 a_0^2 }{\hbar^2 \epsilon_0 c \Gamma_{\rm L}} \frac{1}{3 n^2} \sum_{l m}	\sum_q \sum_{l' m'} \left| \langle n  l m | r^{(q)}/a_0| n' l' m'  \rangle \right|^2
= \frac{2 I e^2 a_0^2}{\hbar^2 \epsilon_0 c \Gamma_{\rm L}}
\frac{1}{3 n^2}	 \sum_{l' l} \left| \langle n' l' \| r/a_0 \| n l \rangle \right|^2 
$.

\subsection{Photoionization cross section}

Using  $R_{nl}^{k l'} =  \int_0^\infty R_{nl}(\rho) \rho R_{k l'}(\rho)  \rho^2 d\rho $,
 the photoionization cross section from 
$nl$ to $k l' $ is given by:
\begin{eqnarray*}
\sigma_{nl }^{k l'  } & =& \frac{4 \pi^2 \omega  a_0 2 Ry  }{3 c } \frac{ \max(l,l')}{2l+1}  \left( {\cal R}_{nl}^{k l'}\right)^2 \\
&= &\frac{4 \pi^2 \alpha a_0^2}{3  }\left(\frac{1}{n^2} + k^2 \right) \frac{ \max(l,l')}{2l+1}  \left( R_{nl}^{k l'}\right)^2\\
&= &\frac{4 \pi^2 \alpha a_0^2}{3  }\left(\frac{1}{n^2} + k^2 \right)	\frac{|\langle k l'  \| r/a_0 \| n l  \rangle| ^2}{2 l +1},
\end{eqnarray*}

It is the cross section assuming that the atoms are evenly distributed among all $2l + 1$
possible $m$ initial states.
So, for a light of given polarization $q$: 
$\sigma_{nl }^{k l'  }  = \frac{1}{2l+1} \sum_m  \sigma_{nl m }^{k l' m'=m+q }$, where
$\sigma_{nl m }^{k l' m'=m+q }$ is
the photoionization cross section from 
$nlm$ to $k l' m'$
given by
\begin{eqnarray*}\sigma_{nl m }^{k l' m' } &=&
3 \left( C_{l m, 1 q=m'-m}^{l' m'} \right)^2 \frac{2l+1}{2l'+1} \sigma_{nl }^{k l'  } \\
&=& 3  \left( C_{l' m', 1 m-m'}^{l m} \right)^2 
\sigma_{nl }^{k l'  } \\
&= & 4 \pi^2 \alpha a_0^2\left(\frac{1}{n^2} + k^2 \right) |\langle k l' m'  | r^{(m'-m)}/a_0 | n l m \rangle| ^2.
\end{eqnarray*}

 $R_{nl}^{k l'} $ is given for $l'=l+1$ and for $l'=l-1$  by:
 \begin{widetext}
\begin{eqnarray}
 R_{nl}^{k l+1} & =& \frac{-i}{4 k (2 l+1)!}
\sqrt{\frac{1}{2} \frac{(n+l)! \prod_{s=1}^{l+1} ( 1 + s^2 k^2)}{(n-l-1)!(1-e^{-\frac{2 \pi }{k}})}
	}	
\left(		
\frac{4 n   }{1+n^2 k^2}\right)^{l+2} 
e^{-\frac{2}{k} \arctan (nk) } 
\left(		
\frac{n - i/k   }{n+i/k}\right)^{n-l-2} 
\label{radial_overlap_continuum_1} \\
& &
\left(
\, _2F_1\left(l+2-i/k,l+1-n;2 l+2;-\frac{4 n\, i/k}{\left(n-i/k\right)^2}\right)-
\left(		
\frac{n - i/k   }{n+i/k}\right)^{2}  \,
_2F_1\left(l-i/k,l+1-n;2 l+2;-\frac{4 n i/k}{\left(n-i/k\right)^2}\right)
\right) \nonumber\\
R_{nl}^{k l-1} & =& \frac{-1}{4  (2 l+1)!}
\sqrt{\frac{1}{2} \frac{(n+l)! \prod_{s=1}^{l-1} ( 1 + s^2 k^2)}{(n-l-1)!(1-e^{-\frac{2 \pi }{k}})}
}	
\left(		
\frac{4 n   }{1+n^2 k^2}\right)^{l+1} 
e^{-\frac{2}{k} \arctan (nk) } 
\left(		
\frac{n - i/k   }{n+i/k}\right)^{n-l-1} 
\label{radial_overlap_continuum_2} \\
& &
\left(
\, _2F_1\left(l-i/k,l+1-n;2 l;-\frac{4 n\, i/k}{\left(n-i/k\right)^2}\right)-
\left(		
\frac{n - i/k   }{n+i/k}\right)^{2}  \,
_2F_1\left(l-i/k,l-1-n;2 l;-\frac{4 n i/k}{\left(n-i/k\right)^2}\right)
\right) \nonumber.
\end{eqnarray}
\end{widetext}

We do not treat here the
continuum-continuum transition (they are given in \cite{gordon1929berechnung}).

\bibliographystyle{unsrt}
\bibliography{2018_bibli_oct_v2_CM}

\end{document}